\documentclass[journal]{IEEEtran}
\ifCLASSINFOpdf
   \usepackage[pdftex]{graphicx}
\else
\fi
\usepackage{subcaption}
\usepackage{booktabs}
\usepackage{hyperref}

\usepackage{xcolor}

\begin{document}
%
\title{Speech Driven Talking Face Generation from a Single Image and an Emotion Condition}
%
%
%

\author{Sefik~Emre~Eskimez,~\IEEEmembership{Member,~IEEE,}
        You~Zhang,~\IEEEmembership{Student~Member,~IEEE,}
        and~Zhiyao~Duan,~\IEEEmembership{Member,~IEEE}
\thanks{S. E. Eskimez was previously with the Department
of Electrical and Computer Engineering, and Y. Zhang and Z. Duan are currently with the Department
of Electrical and Computer Engineering, University of Rochester, Rochester,
NY, 14627 USA e-mail: emreeskimez@gmail.com.}%
}
\maketitle

\begin{abstract}
Visual emotion expression plays an important role in audiovisual speech communication. In this work, we propose a novel approach to rendering visual emotion expression in speech-driven talking face generation. Specifically, we design an end-to-end talking face generation system that takes a speech utterance, a single face image, and a categorical emotion label as input to render a talking face video synchronized with the speech and expressing the conditioned emotion. Objective evaluation on image quality, audiovisual synchronization, and visual emotion expression shows that the proposed system outperforms a state-of-the-art baseline system. Subjective evaluation of visual emotion expression and video realness also demonstrates the superiority of the proposed system. Furthermore, we conduct a human emotion recognition pilot study using generated videos with mismatched emotions among the audio and visual modalities. Results show that humans respond to the visual modality more significantly than the audio modality on this task.
\end{abstract}


%
\IEEEpeerreviewmaketitle

\section{Introduction}

\IEEEPARstart{S}{peech} communication does not solely depend on the acoustic signal. Visual cues, when present, also play a vital role. The presence of visual cues improves speech comprehension \cite{binnie1973bi,bernstein2009auditory,helfer2005role,maddox2015auditory} in noisy environments and for the hard-of-hearing population. Consequently, researchers developed systems that can automatically generate talking faces from the speech in order to provide visual cues when they are not available ~\cite{Chen_2019_CVPR, chung2017you, eskimez2018generating, jamaludin2019syn, song2018talking, vougioukas2019end, yu2019mining, zhou2019talking}. These systems can increase the accessibility of abundantly available audio-only resources for the hearing impaired population and can also increase the quality of human-computer interactions \cite{4637888, 4456693}. They also have broad applications in entertainment, education, and healthcare.

During speech communication, emotion has a direct impact on the transmitted message and can change the meaning drastically~\cite{alpert1963transient}.  
Studies have shown that predicting emotions purely from speech audio is quite difficult for untrained people~\cite{eskimez2016emotion} and that we heavily rely on visual cues in emotion interpretation~\cite{esposito2009perceptual}. Therefore, to make the visual rendering more realistic and to improve speech communication, it is important for automatic talking face generation systems to render visual emotion expressions.  

One approach to emotional talking face generation is to first estimate the expressed emotions from the speech utterance and then render them in the generated talking faces. This approach, however, is limited by the speech emotion recognition accuracy and does not allow independent control of emotion expression in the visual rendering. In this work, we take a different approach: we ignore emotions expressed in the speech audio and condition the talking face generation on an independent emotion variable. This approach provides direct and more flexible control of visual emotion expression and can enable more personalized applications in entertainment, education, and interactive assistive devices. It also provides a powerful tool for behavioral psychologists to conduct emotion-relevant experiments that were not possible before. For example, one can investigate how humans respond to and interact with their conversational partners' emotional expressions by manipulating these emotions in audio and visual modalities independently.

In this work, we propose the first neural network system that generates emotional talking faces from speech conditioned on categorical emotions. The network takes a speech utterance, a reference face image, and a categorical emotion condition as inputs then generates a talking face that is synchronized with the input speech and contains emotional expressions. 
Our main contributions are as follows:
\begin{itemize}
  \item We propose a new talking face generation method that can be conditioned on categorical emotions.
  \item We propose an emotion discriminative loss that classifies rendered visual emotions.
  \item We conduct a pilot study on human emotion perception using talking face videos with mismatched emotions among the audio and visual modalities.
\end{itemize}

The rest of the paper is organized as follows: We first present related work on talking faces in Section~\ref{sec:relwork}. We then describe the proposed method and objective functions in Section~\ref{sec:method}. Then, we present experimentation details, the objective evaluations, and Amazon Mechanical Turk (AMT) subjective evaluations in Section~\ref{sec:exp}. Finally, we conclude the paper in Section~\ref{sec:conc}. Our source code is publicly available\footnote{\url{https://github.com/eeskimez/emotalkingface}}.

\section{Related Work}
\label{sec:relwork}

\subsection{Emotion Models}
In affective computing, researchers leverage emotion models in order to develop automatic systems that can detect emotions. The most utilized emotion models for automatic systems are 1) categorical models 2) dimensional models. Readers are referred to~\cite{scherer2000psychological,ps2017emotion,bourgais2018emotion,cambria2012hourglass,susanto2020hourglass} for more comprehensive coverage of emotion models.

Categorical emotion models assume there are a small number of emotions that are hard-wired to the human brain~\cite{ekman1999basic,ekman2013emotion}. Ekman's model suggested six basic emotion categories: anger, disgust, fear, happiness, neutral, and sadness.

Dimensional models argue that emotions are correlated, and each emotion category can be represented with a combination of values from emotional dimensions~\cite{mehrabian1974approach,russell1980circumplex,lewis2007neural,nicolaou2011continuous,cambria2012hourglass,susanto2020hourglass}. The most famous example is the arousal-valance (AV) dimensions, where each emotion is represented with an arousal axis that determines if the emotion is active or passive and with a valance dimension that determines if the emotion is positive or negative. Another example is the hourglass of emotions~\cite{cambria2012hourglass,susanto2020hourglass}. These models allow a more precise representation of emotions compared to the categorical models.

It should be noted that the selection of an emotion model for an affective computing task highly depends on the availability of the labels.

\subsection{Multimodal Emotion Analysis} \label{subsec:emo_percep}
In this section, we cover some of the key trends in multimodal emotion analysis. Cambria et al.~\cite{cambria2017affective} stated that affective computing and sentiment analysis is an interdisciplinary effort in combining traditional psychological emotion research with machine learning. There are many works on multimodal sentiment analysis~\cite{cambria2013sentic,poria2016fusing,soleymani2017survey, zadeh2018multimodal,chaturvedi2019fuzzy,kaur2019multimodal,susanto2021ten,tzirakis2021end,stappen2021sentiment}. 

Chaturvedi et al.~\cite{chaturvedi2019fuzzy} proposed a fuzzy sentiment classifier to predict mixed sentiment. Ma et al.~\cite{ma2020survey} summarized empathetic dialogue systems, identifying that most of the systems focus on emotion-expressiveness or emotion-awareness. In emotion-expressive systems, emotion labels are employed for the loss calculation. In emotion-aware systems, emotion labels are taken as additional input, which is more similar to our approach. Another important line of research is sentic blending~\cite{cambria2013sentic,susanto2021ten,stappen2021sentiment}. The conceptual and emotional information related to natural language can be defined as semantic and sentic, respectively. The key idea is to fuse many single modality systems that have different time scales and output labels. 

Our emotion study fits well with this research line but is novel since our generated videos with mismatched emotion provide more insights on which modality humans rely on.

\subsection{Emotional Talking Face Generation}
The automatic generation of talking faces from the speech is drawing increasing attention from researchers in recent years. One approach is to first convert the speech input to face landmarks~\cite{chung2017you,suwajanakorn2017synthesizing,song2018talking,karras2017audio,vougioukas2018end,vougioukas2019end,Chen_2019_CVPR,zhou2019talking} and then estimate video frames using the predicted landmarks.
In Suwajanakorn et al.'s two-stage system \cite{suwajanakorn2017synthesizing}, a long short-term memory (LSTM) network first predicts the principal component analysis (PCA) coefficients of face landmarks from speech features, and then retrieves candidate frames from the dataset according to the PCA coefficients, stitching them together. 
However, this system works only for a single speaker. 
Another two-stage system was proposed by Chen et al. \cite{Chen_2019_CVPR}. The system first predicts 68 face landmarks from speech using an LSTM-based network \cite{eskimez2018generating}, and then predicts a few talking face images from the conditioned image and the face landmarks. They also employ a discriminator network to improve image quality.
In another work, Egor et al. \cite{zakharov2019few} proposed a style-based landmark-to-image conversion method using generative adversarial networks (GANs) with a few shots of the target face. This method, however, lacks landmark adaptation methods to solve personality mismatch issues.

Some researchers designed systems that directly map speech features to video frames. Features extracted from the speech often include the Mel-frequency cepstral coefficients (MFCCs), energy, and the first- and second-order temporal derivatives of these features. Gutierrez et al.~\cite{gutierrez2005speech} proposed an integral system that employs the $k$ nearest-neighbor (KNN) algorithm to map the speech dynamics to video frames. The KNN procedure requires a lot of memory and has a long search time, which leads to impracticality in real applications. Some approaches that model the conversion from speech features to the movement of articulators \cite{4130381, 4668497, 1359859}, but they only focus on specific regions of the face, thus generating less natural or expressive animation. Chung et al.~\cite{chung2017you} proposed a convolutional neural network (CNN) that takes as input a face image and speech features and generates a talking face video. The generated video is then sharpened by another CNN, which is trained on pairs of artificially blurred images and their clear originals. Chen et al.~\cite{chen2018lip} proposed another method that predicts video frames of the lip region from speech features and a conditioned lip image. They introduced a GAN loss in addition to the reconstruction loss to sharpen the generated overly smooth video frames.
However, this method is limited to only generating the lip region instead of the entire talking face. Zhou et al.~\cite{zhou2019talking} proposed a GAN-based method that models the whole face and introduced a temporal-GAN loss in addition to the reconstruction loss to improve the temporal dependency across frames. Song et al.~\cite{song2018talking} proposed another method that generates talking faces by using a conditional recurrent adversarial network to improve the realness. Yu et al.~\cite{yu2019mining} adopted optical flow and a self-attention mechanism to capture adjacent and long-range temporal dependencies across video frames.

In addition to the above-mentioned two-stage or speech-feature-driven approaches, there are also end-to-end systems that generate talking faces directly from a conditioned image and the speech signal. Vougioukas et al.~\cite{vougioukas2018end} proposed a temporal-GAN method to generate more realistic image sequences. They further improved their methods with three discriminators~\cite{vougioukas2019end} that focus on improving the realness of video frames, the continuity between generated frames, and the synchronization between audio and visual data. Eskimez et al.~\cite{eskimeze2e} proposed an end-to-end talking face generation system that is robust to noisy speech input. The system contains a frame discriminator to improve image quality and a pair discriminator to improve lip-speech synchronization. They proposed a mouth region mask (MRM) to further improve the lip-speech synchronization and showed that it leads to better alignment than the baselines. 
 
Regarding emotional talking face generation, existing work is somewhat limited. Cosatto et al.~\cite{cosatto2000photo} sample facial details from a database and then project to a 3D head model to allow realistic expressions. However, the generated emotional expressions focus more on the upper part of the face and lack variation for long animations. Karras et al.~\cite{karras2017audio} adopted an end-to-end network to learn a latent representation of emotion states and use the latent code as a control to generate 3D mesh animations. This method effectively discovers emotion variations in the data, but the learned emotion states are difficult to interpret and do not model facial features such as wrinkled eyes and head motion to generate facial expressions. Sadoughi et al.~\cite{Sadoughi2019emo} extended the conditional-GAN-based model to take the target emotion as an input, but this method is limited to generating the lip area instead of the whole face.

Recently, Fang et al.~\cite{fang2021facial} proposed a talking face generation system that takes audio and an image as input. 
This system enforces the generated videos to convey the emotion contained within the speech input. This is different from our method, which allows independent control of the emotion of the generated visual signal from that of the audio input. Also, their generated videos contain a high amount of visual artifacts (e.g., ambiguity, pixel jittering, face deformation) that render them unrealistic.

\subsection{Multimodal Human Emotion Perception}
Emotion perception from auditory and visual stimuli has been examined in recent years. Existing work ~\cite{schirmer2017emotion, douglas2005multimodal, jessen2013role, cowie2009perceiving, busso2004analysis} concludes that different modalities complement each other and that there are also intermodal effects. Cowie~\cite{cowie2009perceiving} showed that perception is sensitive to stimuli from multiple modalities in data from both simulated and natural interactions. Jessen et al.~\cite{jessen2013role} suggested that emotional visual content yields a more reliable prediction of auditory information. Schirmer et al.~\cite{schirmer2017emotion} explored modalities in terms of neural responses and showed that each modality provides a distinct insight, and that multimodal perception converges for holistic emotion recognition.

Most of the existing work was focused on emotionally congruent stimuli from these two modalities; little work examined incongruent stimuli. Tsiourti et al. ~\cite{tsiourti2019multimodal} investigated human responses to emotions expressed by the body and voice of humanoid robots, showing that cross-modal incongruency decreased emotion recognition accuracy. 
Piwek et al.~\cite{piwek2015audiovisual} found that subjects weighted visual cues higher in emotion judgments when presented emotionally incongruent audiovisual clips with happy or angry emotion. However, the visual content was conveyed by point-light displays instead of natural images.
 
\section{Method}
\label{sec:method}
\begin{figure*}[t!]
  \centering
  \includegraphics[width=0.8\linewidth]{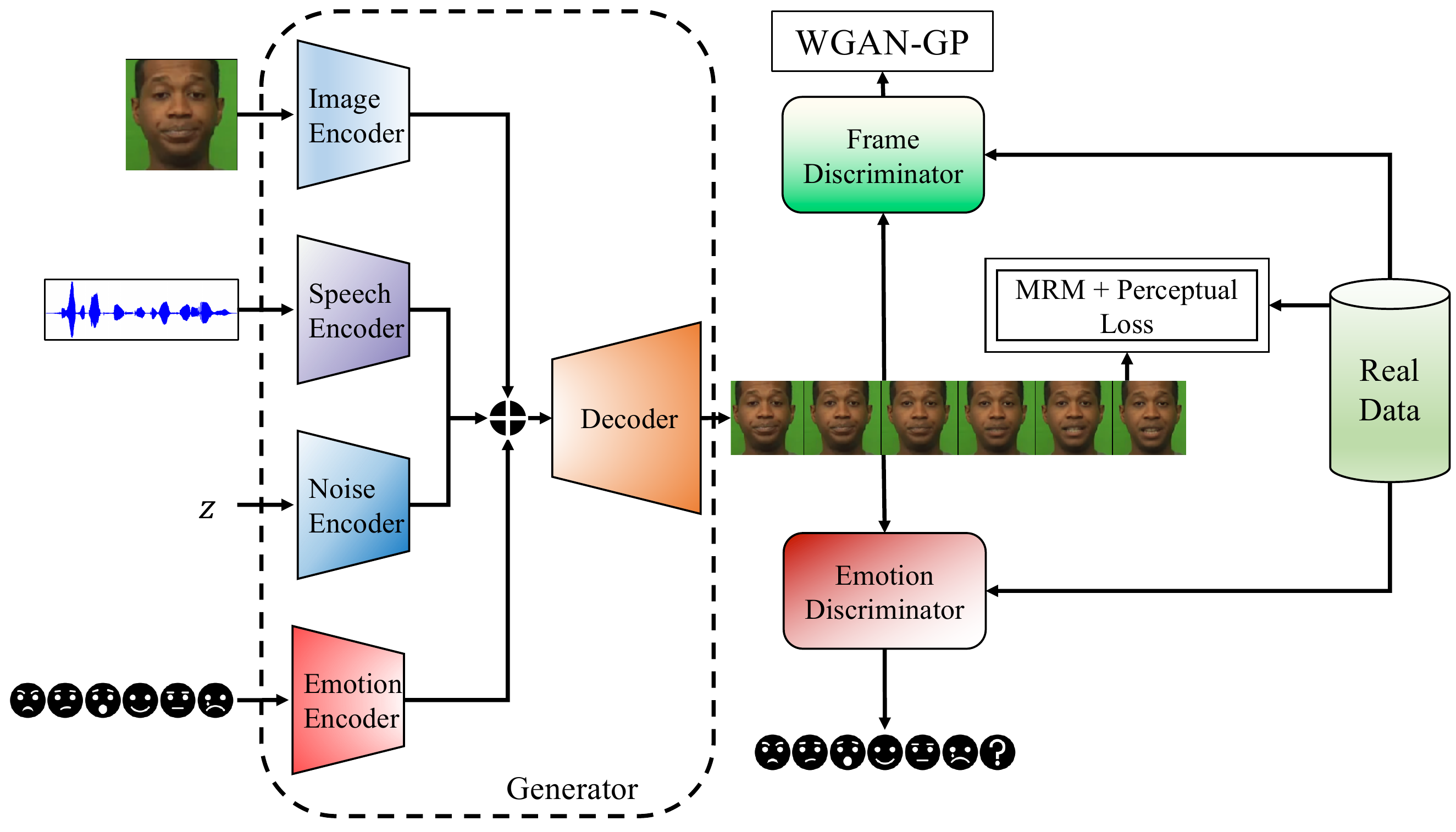}
  \caption{Overview of the proposed neural network system. It accepts a reference image, a speech waveform, a random vector from the standard normal distribution, and a categorical emotion as input, concatenates their embeddings, and generates a talking face video that is synchronized with the input speech and expresses the input emotion. During training, besides the MRM reconstruction loss and perceptual loss, the network employs two discriminative losses: the frame discriminator for image quality and the emotion discriminator for emotion expression.}
  \label{fig:sysoverview}
\end{figure*}

Instead of inferring emotion from the input speech~\cite{karras2017audio, Sadoughi2019emo}, in this work, we propose to use emotions as an input condition to our system. The motivation is to decouple the speech and emotion conditions. This allows us to manipulate emotions during the generation of face videos. Figure~\ref{fig:sysoverview} shows an overview of the system, which employs the GAN framework.
Our generator network architecture is built based on our previous work~\cite{eskimeze2e}, with a modification to accept the emotion condition input. 
For the discriminator networks, we use one discriminator to distinguish between emotions expressed in videos, and another discriminator to distinguish between the real and generated video frames. 

\subsection{Generator}
The generator network contains the following sub-networks: speech, image, noise, and emotion encoders, and a video decoder. 

\subsubsection{Speech Encoder}
The speech encoder processes the input speech waveform and outputs a speech embedding. It follows the original implementation of~\cite{eskimeze2e} without any modification. It contains five convolutional layers with 1-D kernels operating in the time domain. The number of filters, filter sizes, and strides for these layers are as follows: (64, 63, 4), (128, 31, 4), (256, 17, 2), (512, 9, 2), (16, 1, 1), respectively. Each convolutional layer is followed by a LeakyReLU activation with a 0.2 slope. Since our network accepts 8 kHz speech signals, our speech encoder outputs 125 feature vectors per 1 second of speech. We add a context layer after these five convolutional layers to concatenate the past and future speech features. The context layer reduces the 125 time-steps to 25 time-steps by passing only every fifth frame to the next layer. Therefore, our generated videos are in 25 frames-per-seconds (FPS). The output of the context layer is fed to a fully connected layer, followed by two LSTM layers, which output the speech embedding sequence.

\subsubsection{Image Encoder}
The image encoder computes an image embedding from the input condition face image. 
The architecture follows the original implementation without any modification~\cite{eskimeze2e}. It contains six layers of 2-D convolutional layers with the following number of filters, kernel sizes, and down-sampling factors: (64, 3, 2), (128, 3, 2), (256, 3, 2), (512, 3, 2), (512, 3, 2), (512, 4, 1), respectively. A LeakyReLU activation with a 0.2 slope follows each convolutional layer. Note that nearest-neighbor interpolation is used for downsampling rather than using strides. This eliminates the artifacts in the generated images. The final image embeddings and intermediate representations are all passed to the video decoder using U-Net style skip connections~\cite{ronneberger2015u}.

\subsubsection{Emotion Encoder}
The emotion label is first encoded as a one-hot vector and fed into the emotion encoder. The emotion encoder uses a two-layer fully connected (FC) neural network to project the one-hot vector to an emotion embedding. This embedding is replicated for each time step. Again, we use a LeakyReLU activation with a 0.2 slope after every FC layer.

\subsubsection{Noise Encoder}
For each frame of the video, we generate a noise vector drawn from the standard Gaussian distribution. A single-layer LSTM processes this sequence of noise vectors and outputs the noise embedding. This module aims to model the head movements that are not correlated with speech, image, and emotion.  

\subsubsection{Video Decoder}
We modify the video decoder described in~\cite{eskimeze2e} to accept the additional emotion embedding. We concatenate the speech, image, noise, and emotion embeddings and feed them into the decoder. For each time step, the decoder uses convolutional layers to project the embeddings into $4\times4$ images using two FC layers and reshape operations. These $4\times4$ images are concatenated channel-wise with the skip connections coming from the image encoder in the U-Net fashion for the next layers, except for the last layer. The number of filters in each convolutional layer is the same as for the corresponding layer in the image encoder. A LeakyReLU activation with a 0.2 slope follows each convolutional layer, except for the last layer, where instead, hyperbolic tangent activation is used since the images are normalized to have values between -1 to 1.

\subsection{Frame Discriminator}

The frame discriminator aims to improve the image quality of the generated video and to keep the target identity consistent throughout the video. First, we repeat the target image for the number of frames in the input video and concatenate them together. Then, each frame is processed by five layers of 2-D convolutional layers. The number of filters, kernel sizes, and strides of these convolutional layers are as follows: (64, 3, 2), (128, 3, 2), (256, 3, 2), (512, 3, 2), (512, 3, 2), respectively. The output is then flattened and fed into a two-layer FC network, which classifies the frame as fake or real. Each layer is followed by a LeakyReLU activation with a 0.2 slope except for the last layer, where we do not use an activation, since our system employs Wasserstein GAN with a gradient penalty~\cite{gulrajani2017improved}. 

\subsection{Emotion Discriminator}
The emotion discriminator is essentially a video-based emotion classifier, with the inclusion of an additional class for fake videos. It aims to improve the emotional expression generated by our network. The first part of the network follows the same architecture as the frame discriminator: five layers of 2-D convolutional layer followed by two FC layers. We process each frame of the video and feed the resulting sequence into an LSTM layer. The last time step of the output of the LSTM layer is fed into an FC layer that outputs probabilities of the seven classes: six emotions (anger, disgust, fear, happiness, neutral, and sadness) plus the fake class as in~\cite{mariani2018bagan}. 
When we take a training step for the discriminator, we calculate the sparse categorical cross-entropy loss using the emotion label for the real video and the fake label for the generated video. When updating the generator, we calculate the sparse categorical cross-entropy loss using the emotion label we used for generating the video.

\subsection{Objective Functions}\label{sec:loss}

Our system employs multiple objective functions that focus on different aspects of the generated videos: an MRM loss proposed in~\cite{eskimeze2e} to improve mouth-audio synchronization, a perceptual loss to improve image quality, a frame GAN loss for image quality, and an emotion GAN loss for emotion expression.

\subsubsection{Mouth Region Mask (MRM) Loss}
The MRM loss is a weighted L1 reconstruction loss between the generated and ground-truth videos around the mouth region. It uses a 2D Gaussian centered at the mean position of mouth coordinates as the weights.
The intuition of MRM is to manually drive the attention of the network to the mouth region to improve the mouth-audio synchronization.

\subsubsection{Perceptual Loss}
We employ a pre-trained VGG-19 network~\cite{Simonyan15} and calculate intermediate features of the following layers from both the generated and ground-truth videos: 4, 9, 18, 27, and 36. Then, a mean-squared loss between these intermediate features is calculated as the perceptual loss to improve image quality. 

\subsubsection{Frame Discriminator Loss}
To further improve the image quality, especially the sharpness, we use a frame GAN loss calculated by the frame discriminator. Instead of the vanilla GAN loss, we use Wasserstein GAN for more stable training. 

\subsubsection{Emotion Discriminator Loss}
To ensure emotion expression in generated videos, we use an emotion GAN loss calculated by the emotion discriminator, which is a categorical cross-entropy loss using six emotion classes plus a ``fake'' class, similar to~\cite{mariani2018bagan}. In a vanilla GAN discriminator, samples are only classified as real or fake, rather than choosing between multiple emotion classes; if the generator only generates samples from a single emotion class all the time, the vanilla discriminator would still classify them as real. The proposed discriminator, on the other hand, incorporates multi-class classification losses and mitigates this issue of mode collapse for a multi-class generation. 

The full objective function for the generator step is as follows:
\begin{equation}\label{cost:gen}
  \begin{array}{r c l}
  	J_{GEN} = \alpha L_{1}^{MRM} + \beta L_{2}^{Perceptual} +  \gamma J_{FD} +  \delta J_{ED} 
  \end{array},
\end{equation}	
where $J_{GEN}$ is the generator loss, $L_{1}^{MRM}$ is the MRM loss, $L_{2}^{Perceptual}$ is the perceptual loss, $J_{FD}$ is the frame GAN loss, $J_{ED}$ is the emotion GAN loss, and $\alpha, \beta, \gamma, \delta$ are the respective weights of each component. 

\section{Experiments}
\label{sec:exp}
In this section, we describe the data used in experiments, the hyper-parameters of the neural networks, and the objective and subjective evaluation procedure. We choose the temporal GAN approach described in~\cite{vougioukas2019end} as our baseline since it is the closest to our method. We use the pre-trained model and inference code provided by the authors to generate baseline videos. Although it cannot control the emotions through a conditioned input, it can generate emotional expressions that are inferred from the speech.

\subsection{Dataset}
We used the Crowd-sourced Emotional Multimodal Actors Dataset (CREMA-D) dataset~\cite{cao2014crema}. It contains video clips of 91 actors (48 male and 43 female) expressing six categorical emotions: anger, disgust, fear, happiness, neutral, and sadness. The age range of the actors is between 20 to 74. Each video clip shows one actor speaking a sentence from a set of 12 sentences and simulating one of the emotion categories. The image resolution of the provided videos is 480x360, and the sampling rate is 30 frames per second (FPS). The audio is sampled at 44.1 kHz. We downsampled the video to 25 FPS and the audio to 8 kHz.
We followed the same train (70\%), validation (15\%), and test (15\%) splits as~\cite{vougioukas2019end}. We used the same files for these splits to ensure a fair comparison. During testing, the speech utterance, the conditioned emotion, and the conditioned image input to the generator network for each generation are all from the same ground-truth video, where the condition image is the first frame of the video. 

As the same actor's face in different videos may be at different spatial locations, for easing the training, we need to align them across videos. For alignment, first, we choose a template image for the actor where the face is symmetrical. We extracted face landmarks from this template image as the template landmarks. Then, for each video of the actor, we estimated the similarity transform parameters between the template landmarks and extracted landmarks of the first frame using three points: the temporal mean points of the left eye, the right eye, and the nose. Note that we only took the first frame of each video to estimate the transformation, and used it to align the remaining frames to the template image. In this way, the faces in the resulting videos start from the same spatial location but can wander off to different parts of the scene. This allows us to model the natural head movements in addition to facial expressions.

During training, we randomly augmented the data using the Albumentations library~\cite{albumentations} to improve the generalization capability of our network. The data augmentation includes randomly changing brightness, contrast, gamma, hue, saturation, and value. In addition, our algorithm includes contrast limited adaptive histogram equalization, adding random Gaussian noise to the image, and shuffling the channels, and shifting RGB values for each channel. 

\subsection{Implementation Details}
To initialize our network, we trained it from scratch using only MRM and perceptual losses for 100k iterations. Then, we trained it for another 100k iteration using the full objective function. We used Adam optimizer for all networks with $\beta_{1}=0.5, \beta_{2}=0.99$. The learning rate for the generator was 1e-4 during the initialization and 1e-5 during the GAN training. Both discriminators' learning rates were 1e-4. The constants $\alpha, \beta, \gamma, and \delta$ mentioned in Section~\ref{sec:loss} were 100, 1, 0.01, and 0.001, respectively. The weight for the gradient penalty when training the frame discriminator was 10. All images were normalized between the -1 to 1 value range. During initialization, the mini-batch size was set to 8, and during GAN training, it was set to 4. The number of frames per sample was set to 32. The training took approximately one week using a GTX 1080 TI GPU. For the baseline method, we use the pre-trained model (trained with the CREMA-D dataset) provided by the authors. 

\subsection{Objective Evaluation}
We evaluated the image quality of the generated videos using Peak SNR (PSNR) and Structural Similarity (SSIM)~\cite{wang2004image} between the generated video frames and the ground-truth video frames. To measure the audiovisual synchronization, we used the normalized landmarks distance (NLMD)~\cite{eskimeze2e} between landmarks extracted from the generated and ground-truth video frames. 

\begin{table}
  \centering
  \caption{Objective evaluation results for the baseline and our proposed method. For PSNR and SSIM, higher values are better; for NLMD, lower values are better.}
  \label{tab:results}
  \begin{tabular}{lccc}
    \toprule
    Method&PSNR&SSIM&NLMD\\
    \midrule
    Baseline~\cite{vougioukas2019end} & 29.64 & 0.82& 0.124\\
    Proposed  & \textbf{30.91} & \textbf{0.85}& \textbf{0.113}\\
  \bottomrule
\end{tabular}
\end{table}

\begin{figure}[ht]
  \centering
  \includegraphics[width=\linewidth]{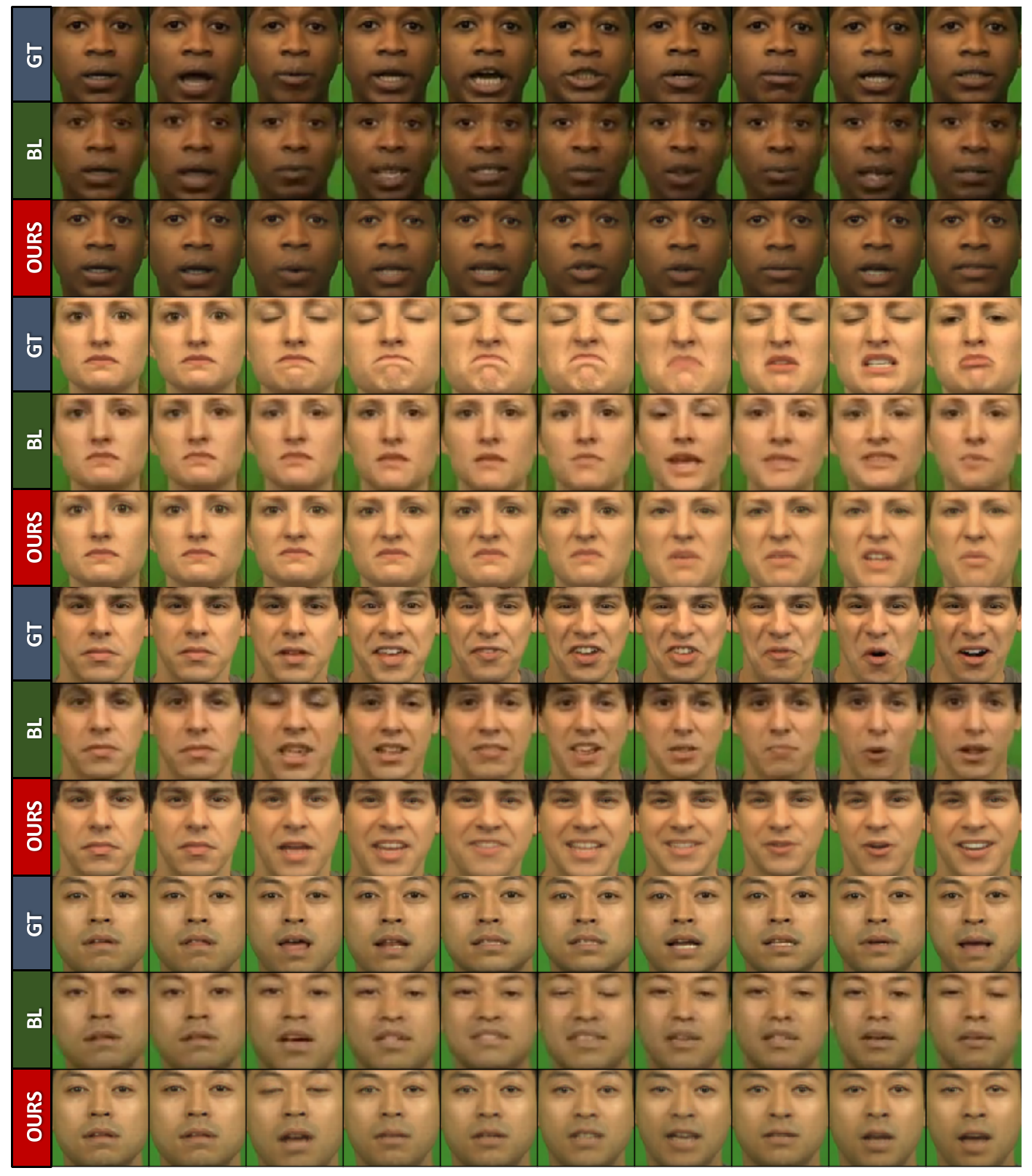}
  \caption{Four examples comparing spatially aligned and cropped videos of the ground-truth (GT), baseline (BL) and proposed approach (OURS) for objective evaluation. Every fifth frame is shown for each video.}
  \label{fig:comp}
\end{figure}

\begin{figure*}[ht]
  \centering
  \includegraphics[width=\linewidth]{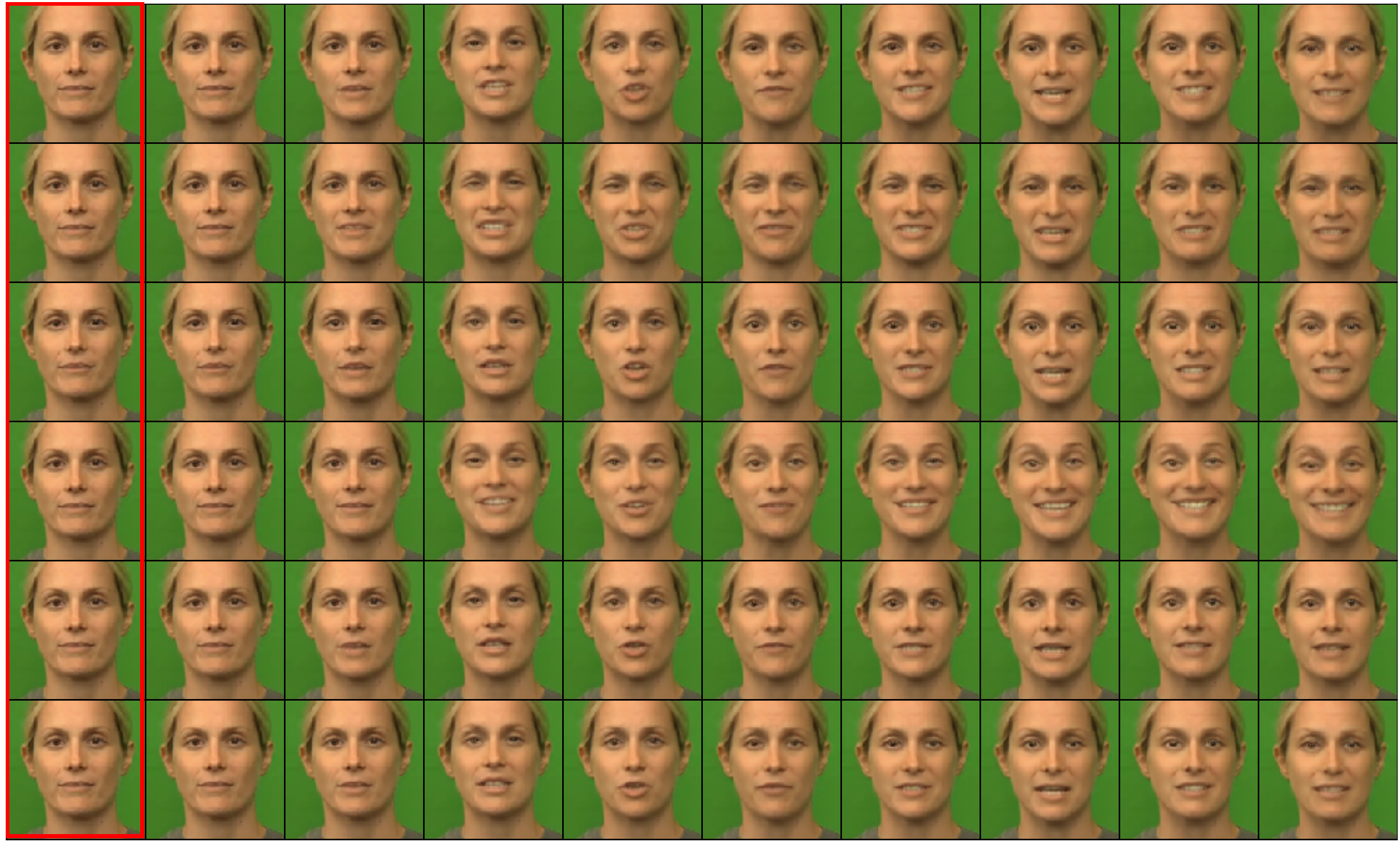}
  \caption{Frames of different talking face videos generated (in different rows) using the same face image (the first column) and speech utterance but different emotion conditions (from top to bottom: anger, disgust, fear, happiness, neutral, and sadness). One frame is shown for every 0.2 seconds.}
  \label{fig:emoexamples}
\end{figure*}

The baseline method generates 96x128 images, while our method yields 128x128 images. In other words, the foreground/background ratio differs in the generated videos. To ensure a fair comparison, we aligned the ground-truth, baseline, and proposed videos into a template image and cropped them into the same size using similarity transformation. Figure~\ref{fig:comp} shows the aligned videos, and Figure~\ref{fig:emoexamples} shows example videos generated from the same condition image and speech, but different emotion conditions.

Table~\ref{tab:results} shows the objective evaluation results of the baseline and our proposed methods. It can be seen that our method outperforms the baseline on all of the three metrics. We believe that perceptual loss is responsible for the improvement in image quality (PSNR and SSIM). For audiovisual synchronization (NLMD), even though our method does not use a discriminator paired with a synchronization loss as in~\cite{vougioukas2019end}, the improvement is as high as 8.9\%, showing the effectiveness of the MRM loss.

\subsubsection{Video-based Emotion Classification} \label{subsubsection:video_emo_class}
In order to validate the emotional expression in the generated videos, we trained a video-based emotion recognition network using the CREMA-D train set. This network uses the same architecture as the emotion discriminator in Figure~\ref{fig:sysoverview}. We then classified the emotions within the ground-truth videos and our generated videos of the test set. The results are shown in Table~\ref{tab:emo_results}. The 6-class emotion classification accuracy on the ground-truth videos is 62.71\%, which is comparable with \cite{ristea2019emotion}, suggesting the validity of the video-based emotion classifier. The accuracy and F1-Score on the generated videos are slightly higher, even though the classifier was not trained on generated videos. This suggests that emotions are well expressed, and slightly exaggerated perhaps, in the generated videos.

\begin{table}
    \centering
  \caption{Video-based emotion classification results for the ground-truth and our generated videos of the test set are shown.}
  \label{tab:emo_results}
  \begin{tabular}{lcc}
    \toprule
    Data&Accuracy&F1-Score\\
    \midrule
    Ground-truth & 62.71 & 62.39\\
    Generated  & 65.67 & 66.65\\
  \bottomrule
\end{tabular}
\end{table}

\begin{figure}[ht]
\centering
\begin{subfigure}[t]{0.8\columnwidth}
  \centering
  \includegraphics[width=\columnwidth]{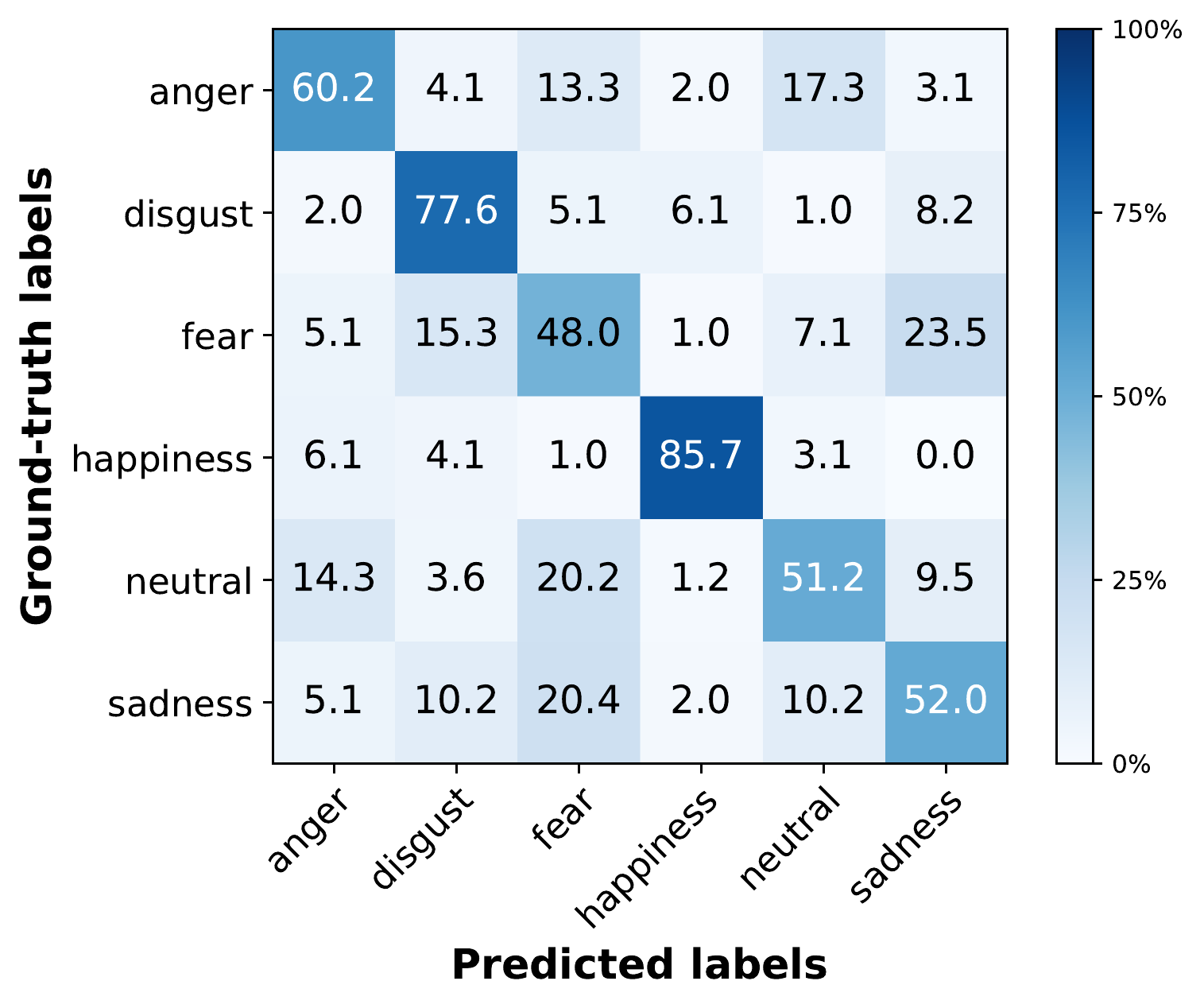}  
  \label{fig:conf_gt}
\end{subfigure}
~
\begin{subfigure}[t]{0.8\columnwidth}
  \centering
  \includegraphics[width=\columnwidth]{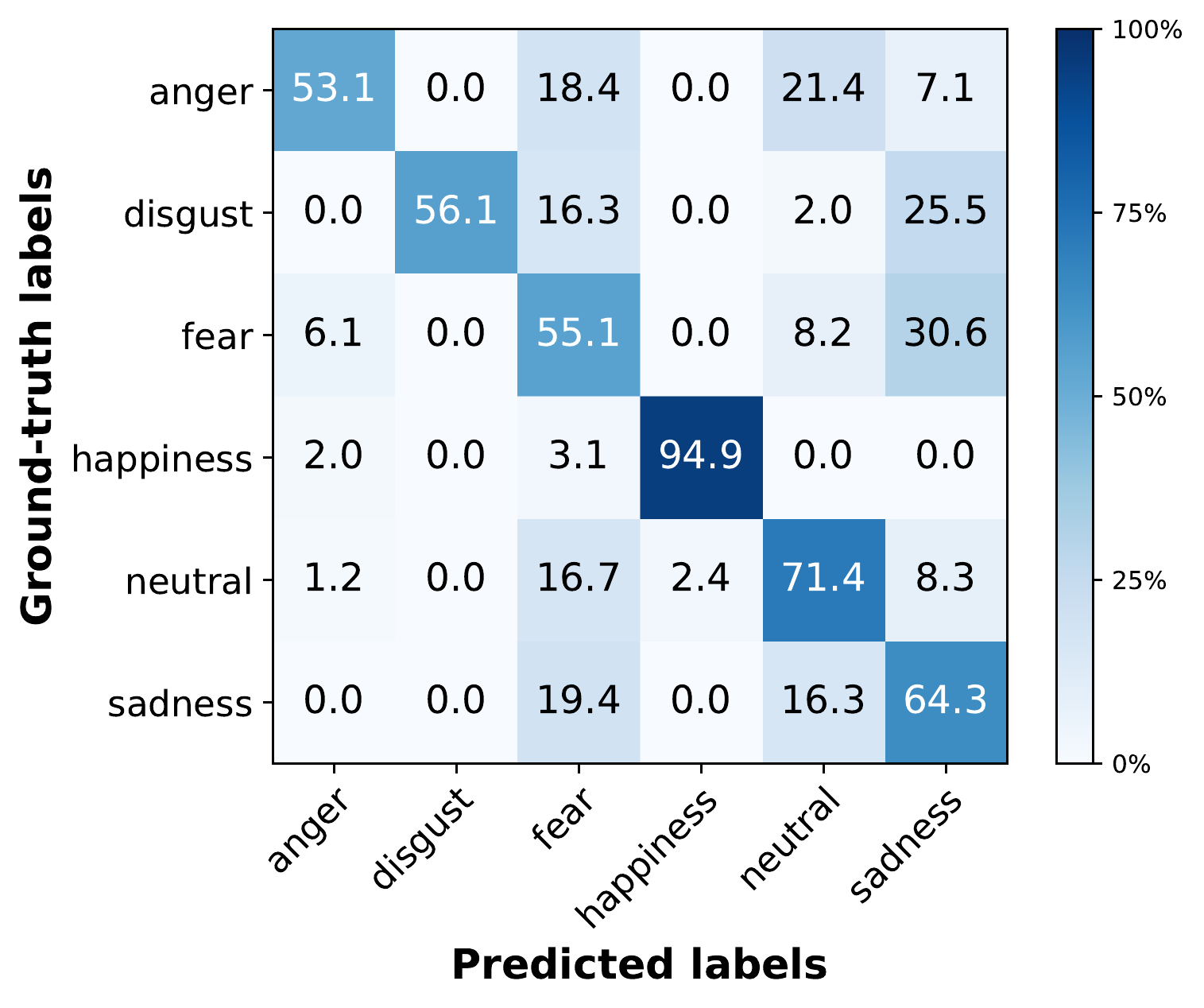}  
  \label{fig:conf_pd}
\end{subfigure}
\caption{Confusion matrices of video-based emotion classification on the ground-truth (above) and generated (below) videos using the proposed talking face generation system on the test dataset of CREMA-D. Each row sums to 100\%. 
}
\label{fig:conf_obj}
\end{figure}

We further show the confusion matrices of these two classification results in Figure~\ref{fig:conf_obj}. We observe similar patterns. First, they both have a strong diagonal. In particular, happiness is the easiest emotion to classify. This may be because happiness often contains smiling that is distinctive from other facial expressions, allowing the classifier as well as our generation system to capture it clearly. Second, some emotions are commonly confused with each other, such as fear and sadness. 
On the other hand, there are also differences in these confusion matrices. In particular, in the ground-truth videos, both fear and sadness are often misclassified as disgust, while in the generated videos, no other emotions are misclassified as disgust. 
Overall, the similarities outweigh the differences, showing that the emotional expressions in the generated videos resemble those of the ground truth. \footnote{For video samples, please visit the project webpage: \url{http://www2.ece.rochester.edu/projects/air/projects/tfaceemo.html}}

\subsection{Subjective Evaluation}

\subsubsection{Research Questions} 

We design our subjective evaluation to investigate the following research questions: 1) Is our model effective in expressing emotions for video rendering? 2) How real are the generated videos of our model? 3) Which modality do people primarily rely on to perceive emotions? We conduct our evaluation on AMT.

\begin{figure*}[ht]
\begin{subfigure}{.33\textwidth}
  \centering
  \includegraphics[width=\linewidth]{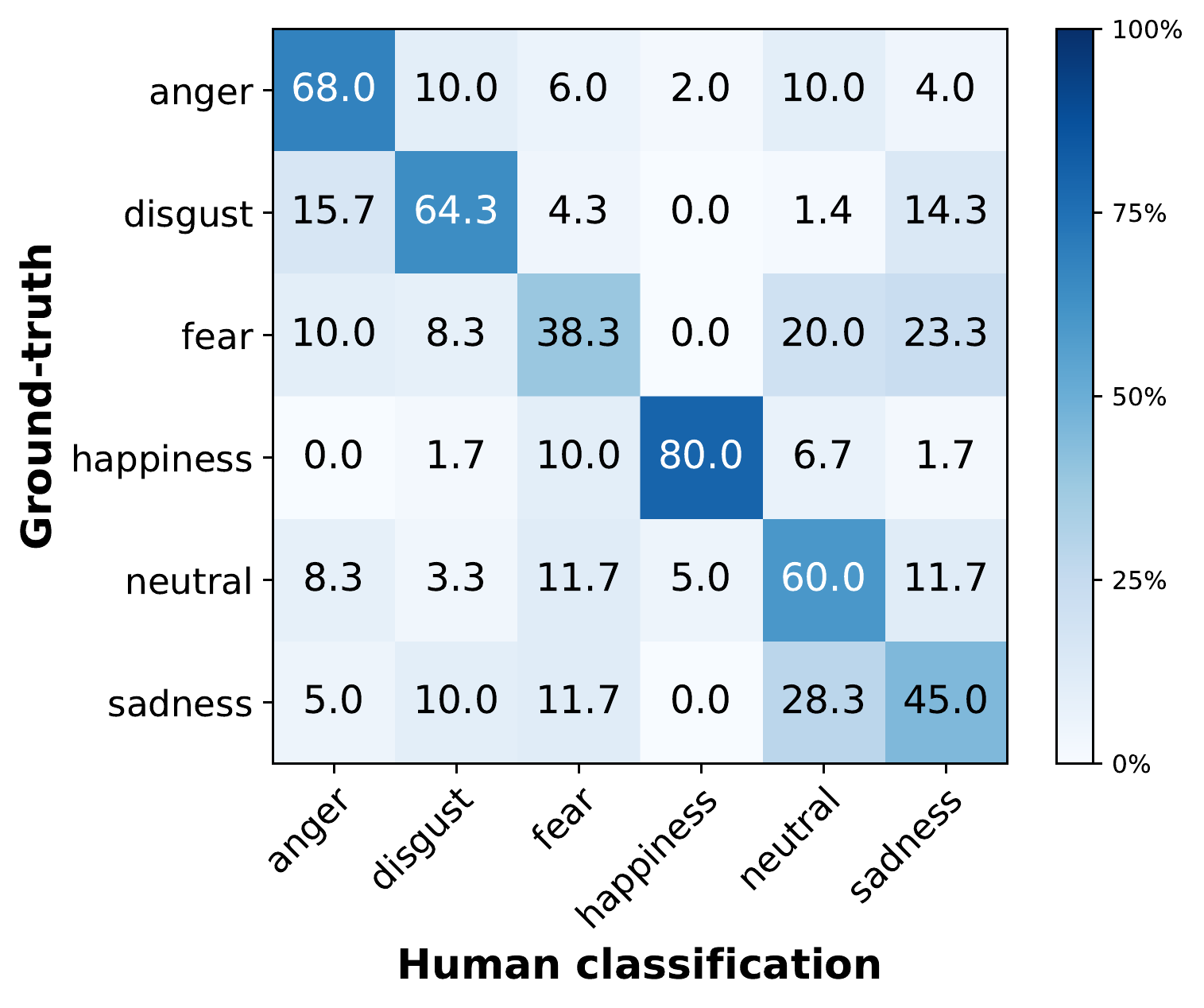}  
  \label{fig:conf_gt_amt}
\end{subfigure}
~
\begin{subfigure}{.33\textwidth}
  \centering
  \includegraphics[width=\linewidth]{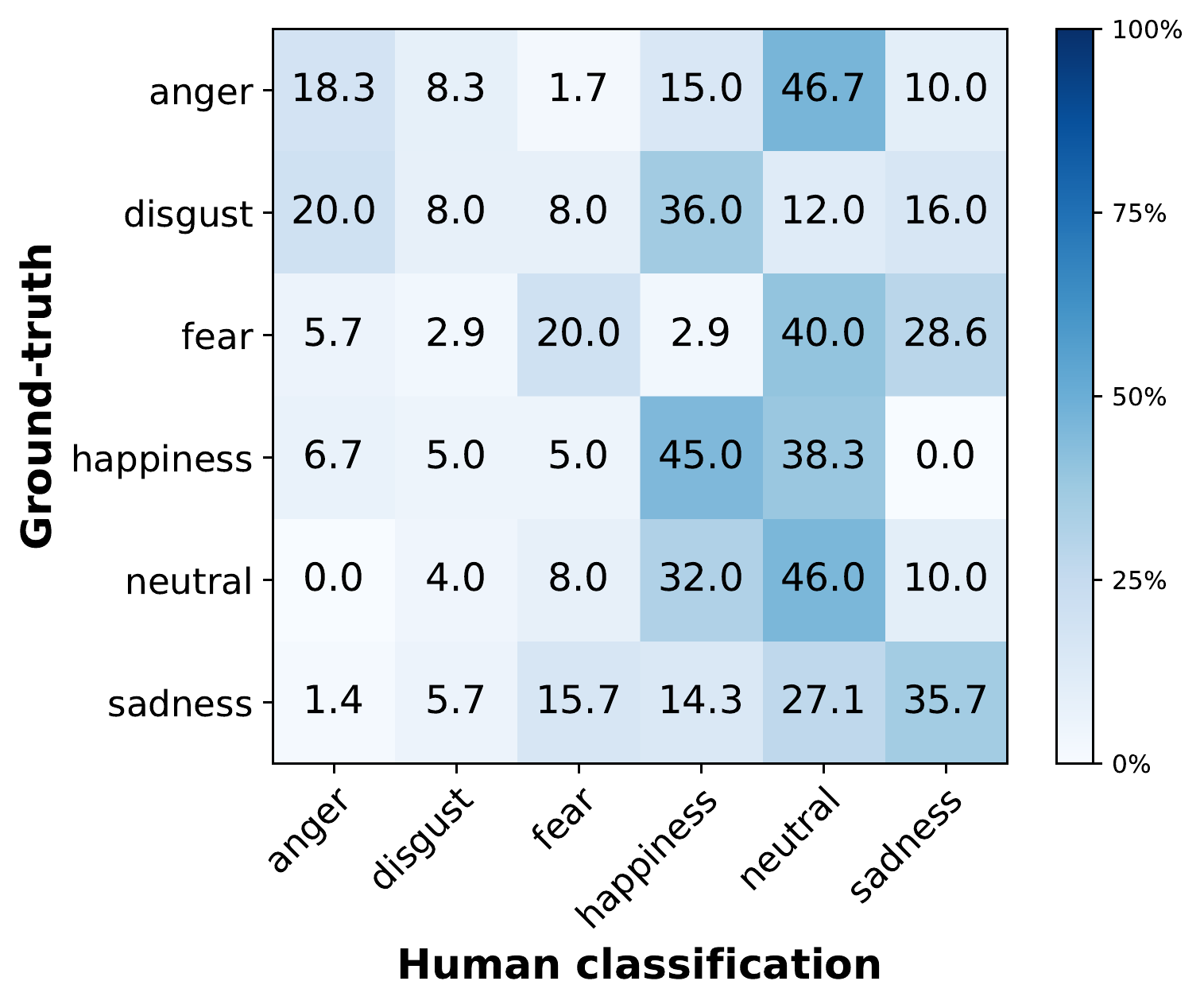}  
  \label{fig:conf_bl_amt}
\end{subfigure}
~
\begin{subfigure}{.33\textwidth}
  \centering
  \includegraphics[width=\linewidth]{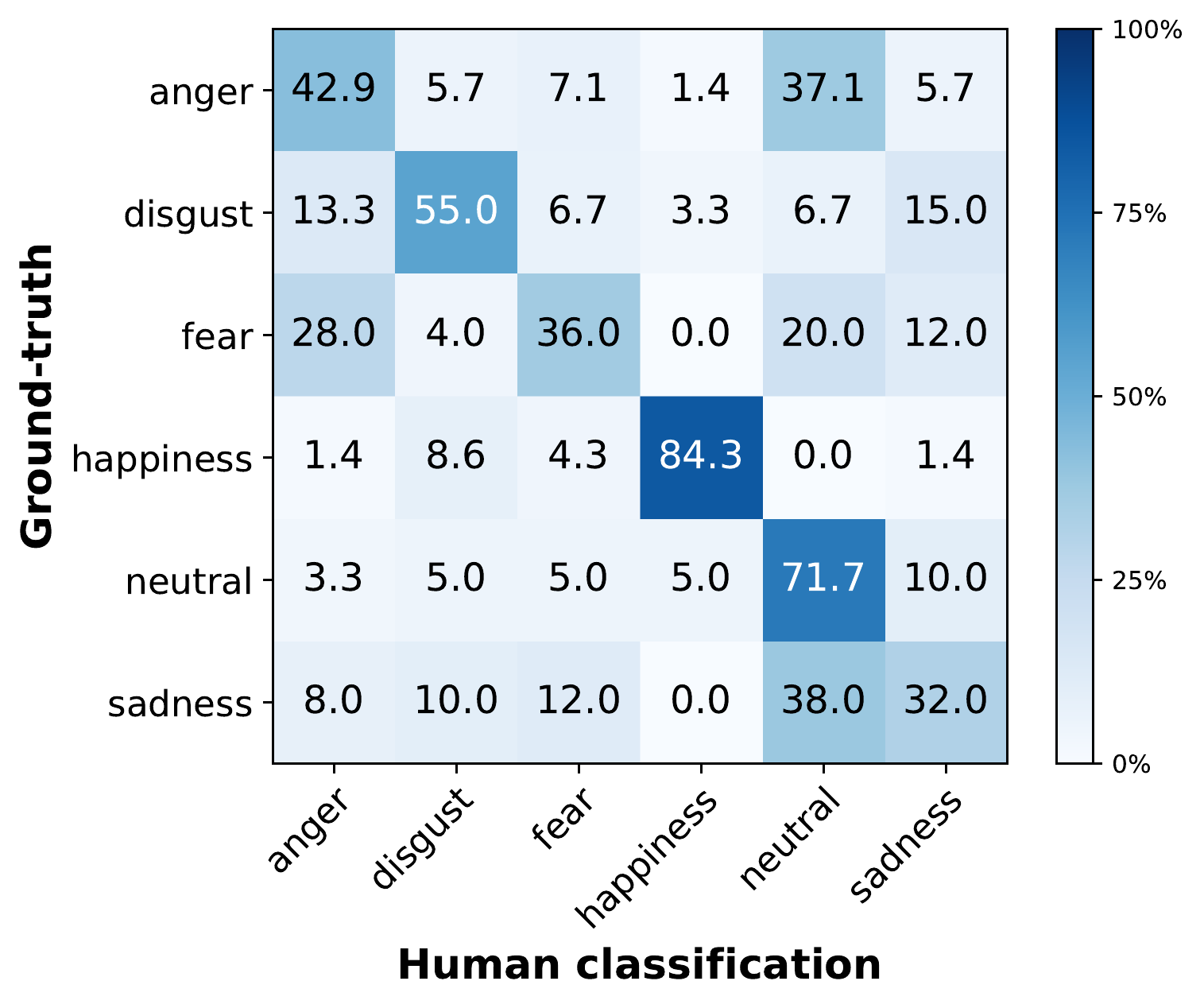}  
  \label{fig:conf_pd_amt}
\end{subfigure}
\caption{Confusion matrices of human emotion classification in Task 1 on ground-truth videos (left), baseline generated videos (middle) and our generated videos (right).}
\label{fig:conf_amt}
\end{figure*}

\subsubsection{Experimental Setup}
Our AMT study consists of two Human Intelligence Tasks (HIT). For the first task, we randomly presented subjects generated and ground-truth videos and asked them to rate the realness and provide suggestions for making the videos more real. We also asked subjects to assign an emotion label to each video. 
This task aimed to answer the first and second research questions. For the second task, we generated videos that contain mismatched emotions in the audio and visual modalities. We asked subjects to assign one or two emotion labels to these videos. By doing so, we aimed to answer the third research question.

\textbf{Task 1 - Emotion Classification and Realness Evaluation.}
In the first task, we pooled videos taken from ground-truth videos of the test set of the CREMA-D dataset and generated videos from the baseline and our models. For the baseline system, each video was generated from the speech recording and the first frame of the ground-truth video, while for the proposed system, each video was generated from the speech recording and the first frame of the ground-truth video, as well as the ground-truth emotion condition. We downsampled the ground-truth and baseline videos to 25 FPS to make them consistent with our generated ones. As described earlier, our method generates talking faces in $128\times128$ image size, while the baseline method generates videos in $96\times128$ image size. If we were to set the ground-truth videos to any of the two sizes, the subjects might be negatively biased toward the generated videos with the other size. To avoid this potential problem, we aligned the ground-truth videos with template faces of both sizes and obtained two sets of ground truth videos. 

We released six batches of videos in total. For each batch, we randomly selected five videos from the two sets of ground-truth videos, five from the baseline videos, and five from our generated videos. One video from each category was repeated to check the consistency of the subjects' answers. Therefore, there were in total 18 videos in each batch. The videos in each batch were randomly shuffled. Across all of the six batches, the total number of videos for each emotion category was equal.

We recruited a total of 60 valid subjects (i.e., 10 for each batch) from AMT. The subjects were required to be located in the United States and to have a lifetime HIT approval rate higher than 95\%. To encourage the subjects to treat the experiments more seriously, we made a bonus payment based on the subject's performance. Subjects were informed about the bonus payment before they started the experiments.

Subjects were informed that some of the 18 videos were recordings of real people, while other videos were rendered by artificial intelligence (AI) based on a single face image of one person and a speech recording of another person. 
Before presenting the 18 videos, we also presented two example videos of real recordings for each emotion, each in the two image sizes ($128\times128$ and $96\times128$), to familiarize the subjects with these emotional expressions. These example emotions were ordered in alphabetical order. 

 We then asked subjects the following three questions for each video: \textit{1) Which emotion is primarily expressed by the person?} This question is a multiple-choice question, and the subjects were asked to select one from the six emotions.
\textit{2) How realistic is the video?} The subjects can choose from \textit{Definitely real}, \textit{Somewhat real}, \textit{Neutral}, \textit{Somewhat unreal}, and \textit{Definitely unreal}. \textit{3) Which aspect(s) can be improved to make the video more real?} This question is a checkbox question, and the subjects could choose more than one aspect. The choices are \textit{None}, \textit{Image quality}, \textit{Lip synchronization}, \textit{Head movement}, and \textit{Other}.

After receiving a survey, we checked its completeness and the consistency of answers to the nine questions of the three repeated videos. We rejected a total of 11 incomplete surveys and those that did not meet the consistency requirement. We then recruited other subjects until we collected 60 valid surveys. For our answer consistency requirement, an answer was considered inconsistent from the previous answer, if 1) the emotion classification was different for the first question, 2) the realness rating differed more than one level for the second question, or 3) the aspect selection differed for more than one options. Among the nine repeated questions, if more than five answers were inconsistent, then the entire survey was rejected.

\textbf{Task 2 - Emotion Perception of Videos with Mismatched Emotions.}
As described in Section~\ref{subsec:emo_percep}, little work on human emotion perception used emotionally incongruent stimuli between the audio and visual modalities, and among these works, none used videos of humans as stimuli. Our emotional talking face generation system makes it possible to investigate human emotion perception from emotionally incongruent stimuli present in human speaking videos.

In the second task, we presented generated videos from our proposed system based on a face image, a speech recording, and an emotion condition. Both the face image and the speech recording were taken from a ground-truth video in the test set of the CREMA-D dataset. As a result, the speech recording conveyed a certain emotion. The emotion condition input, however, was not necessarily the same as the speech emotion, allowing for the possibility of generating videos with mismatched emotions between the audio and visual modalities. As there are six emotions in the dataset, there were 36 emotion pairs and 30 of them were mismatched. We generated 2 videos for each of the 36 pairs, shuffled them, and split them evenly into six batches. We also repeated two videos in each batch to check the answer consistency. Therefore, there were a total of 14 videos in each batch.

We recruited a total of 60 subjects (i.e., 10 for each batch) from AMT, with the same requirements as Task 1. We rejected a total of 4 incomplete surveys and those who had more than two inconsistent answers among the four repeated questions and recruited other subjects until we collected 60 valid surveys. The participants who completed Task 1 could not see this task from the AMT platform. The same bonus mechanism in Task 1 was applied to Task 2.
In the survey, subjects were notified that all of the videos were AI rendered.
Before presenting the generated videos, the subjects were also presented two example ground-truth videos for each emotion, only in the image size of $128\times128$, to familiarize them with the emotions. They were asked the following two multiple-choice questions for each video: \textit{1) Which is the primary emotion expressed by the person?} The subjects could select one of the six emotions.
\textit{2) Which is the secondary emotion expressed by the person?} The subjects could select one of the six emotions and a \textit{None} option if they only perceived the primary emotion. 

\subsubsection{Experimental Results}

\textbf{Task 1 - Emotion Classification.}
The confusion matrices of subjective emotion classification for ground-truth videos, baseline generated videos, and our generated videos are shown in Figure~\ref{fig:conf_amt}. Our videos yield a more diagonal confusion matrix compared with the baseline videos and result in patterns similar to those produced from the ground-truth videos. Specifically, subjects are more likely to classify the emotions in the baseline videos as \textit{neutral}, while this happens much less frequently for our generated videos. This shows the power of the emotion condition input that our method utilizes. The overall classification accuracy is 59.2\% (ground-truth), 28.9\% (baseline), 55.3\% (ours), respectively, demonstrating the efficacy in expressing emotions of our proposed emotional talking face generation system. It must be noted that the baseline system infers emotion from the speech input instead of taking the emotion condition as input. As emotion recognition from the speech is itself a challenging task, errors in this stage naturally influence visual emotion expression in the generated videos. Therefore, poor performance from the baseline system is expected. Interestingly, the 59.2\% human emotion classification accuracy on ground-truth videos is slightly lower than that of our emotional classifier in Section~\ref{subsubsection:video_emo_class}, showing the challenge of visual speech emotion classification for humans. This observation is similar to a speech emotion classification observation in~\cite{eskimez2016emotion}. 

\begin{figure}[ht]
  \centering
  \includegraphics[width=0.6\columnwidth]{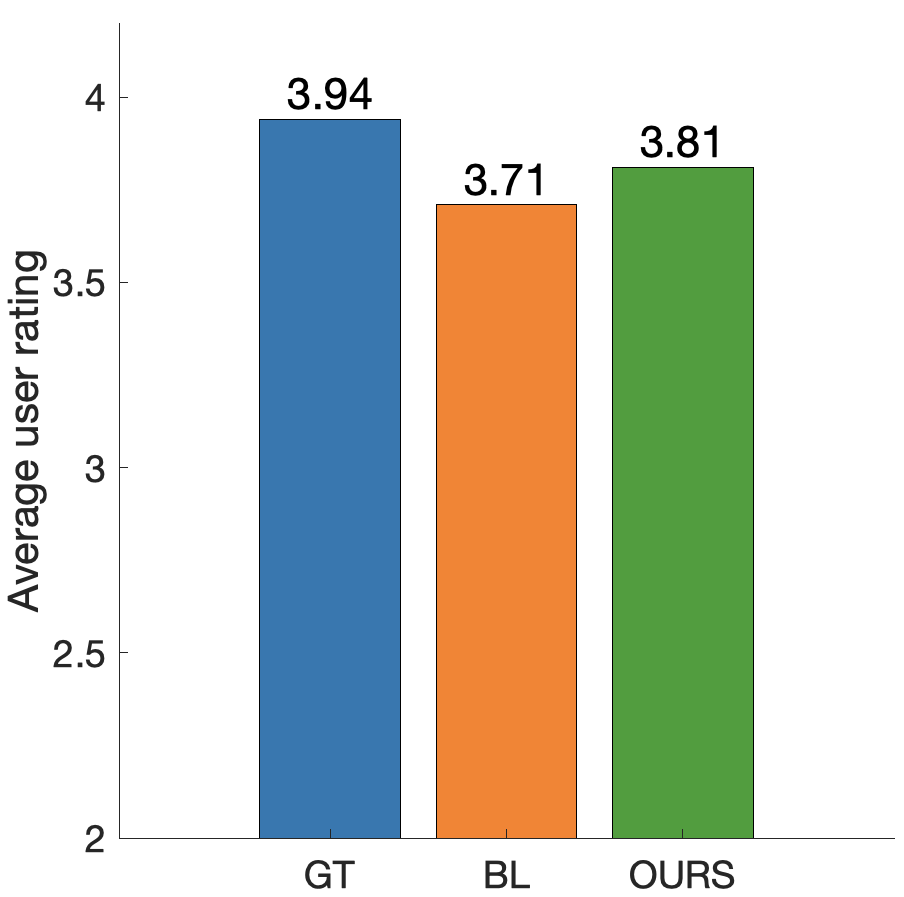}
\caption{User ratings on the realness of ground-truth (GT), baseline generated (BL) and our generated (OURS) videos.}
\label{fig:AMT_rate}
\end{figure}

\textbf{Task 1 - Realness Evaluation.}
For the realness question, the five options are mapped to a scale from 1 to 5, where "definitely real" corresponds to 5 and "definitely unreal" corresponds to 1. The result is shown in Figure~\ref{fig:AMT_rate}. The average rating across all videos and subjects is 3.94, 3.71, and 3.81 for ground-truth, baseline, and our videos, respectively. This suggests that our generated videos are slightly more realistic than the baseline videos, yet they are still not as realistic as the ground-truth videos. Interestingly, even the ground-truth videos only received an average rating close to 4 (somewhat real). We think that this might be due to the relatively lower image resolution than what the subjects typically see in their daily life. This might also because the generated videos (especially OURS) are quite realistic, lowering the subjects' confidence in rating the ground-truth videos.
A Wilcoxon signed-rank test~\cite{cureton1967normal} shows that the median difference between our ratings and the baseline ratings is statistically significantly greater than zero, at the significance level of 0.05 ($p=0.048$).

\begin{figure}[ht]
  \centering
  \includegraphics[width=\columnwidth]{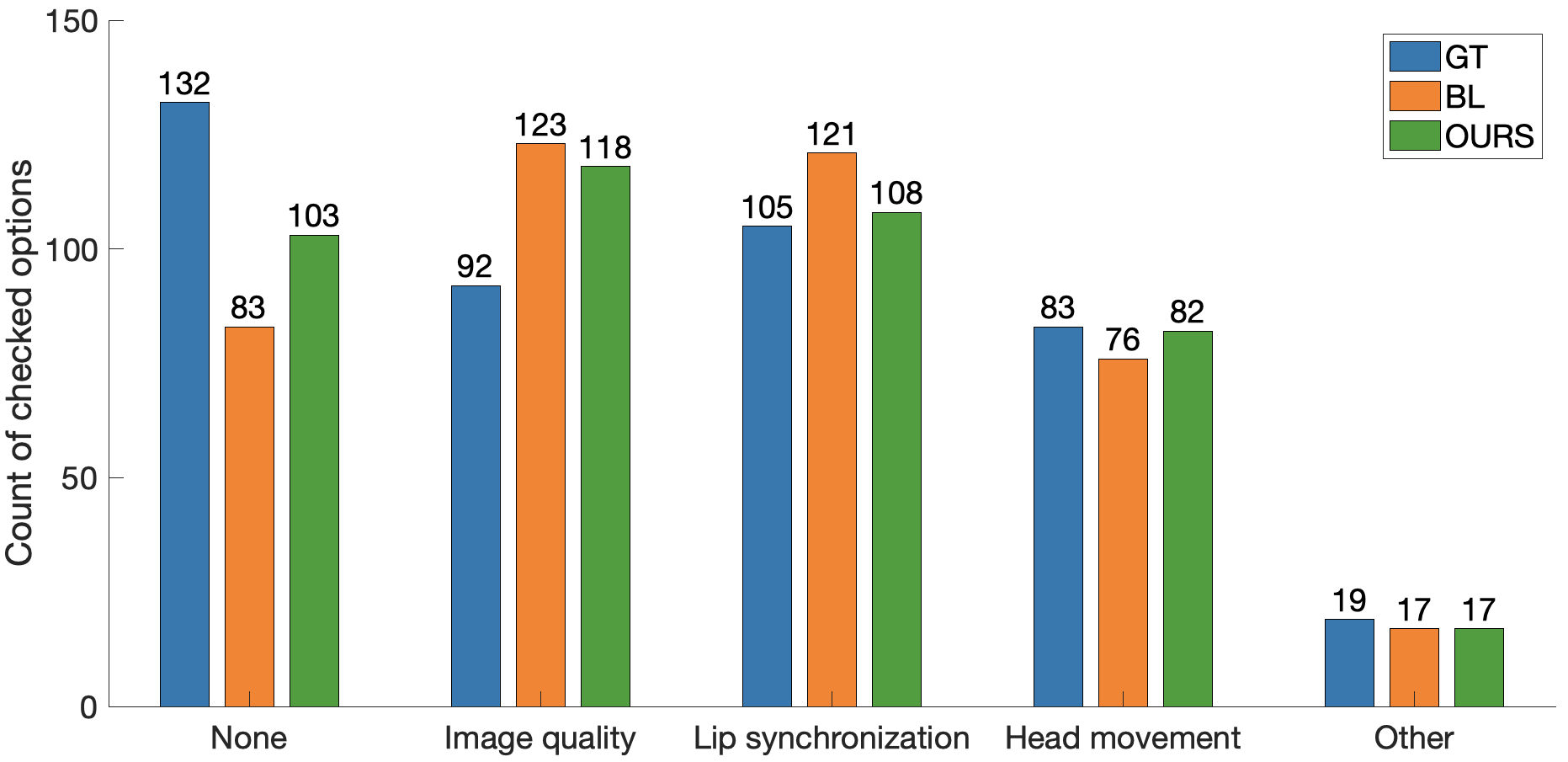}  
\caption{Total count of chosen aspects for realness improvement of videos.}
\label{fig:AMT_barplot}
\end{figure}

Figure~\ref{fig:AMT_barplot} shows the histograms of aspects suggested by the subjects to improve the realness of the videos. Consistent with the realness question, ground-truth videos received the most ``none'' votes, while 
our generated videos received the second most and 
the baseline received the least
. The total count of votes for the four aspects to improve (``image quality'', ``lip synchronization'', ``head movement'', ``other'') is 299 (ground-truth), 337 (baseline) and 325 (ours), respectively. Among the detailed aspects, the baseline videos received the most votes on ``image quality'' and ``lip synchronization''; but it also received the least votes on ``head movement''
. This might be due to the fact that the baseline method is trained with 30 FPS videos and adopted a sequence discriminator to render head movements. On the other hand, our generated videos performed similarly to ground-truth ones on ``lip synchronization'' and ``head movement'', suggesting the effectiveness of our proposed MRM loss. Nevertheless, the ``image quality'' of our generated videos is considered to need more improvement than the ground-truth videos.

\begin{figure}[ht]
\centering
\begin{subfigure}[t]{0.8\columnwidth}
  \centering
  \includegraphics[width=\linewidth]{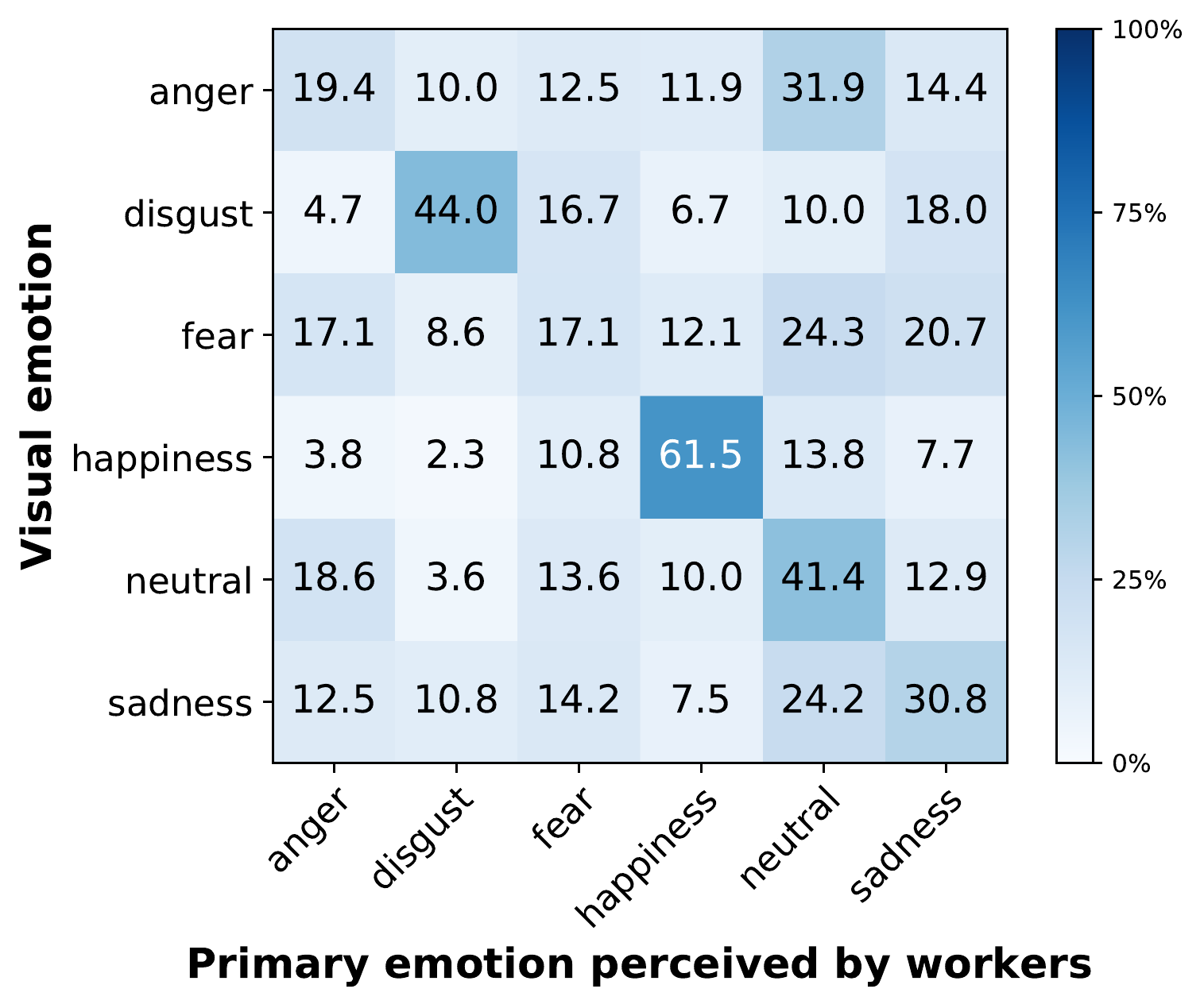}  
  \label{fig:video_primary}
\end{subfigure}
~
\begin{subfigure}[t]{0.8\columnwidth}
  \centering
  \includegraphics[width=\linewidth]{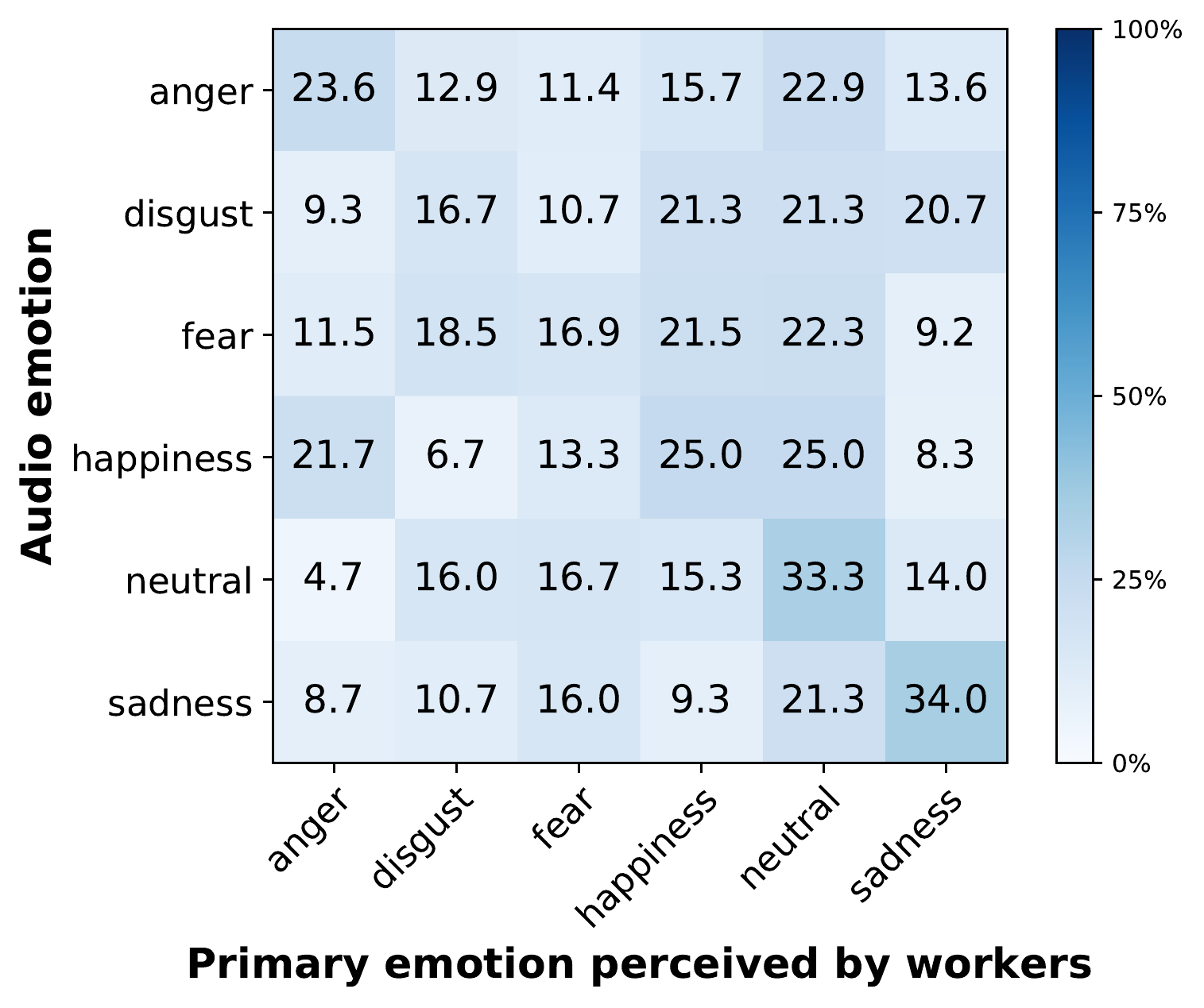}  
  \label{fig:audio_primary}
\end{subfigure}
\caption{Confusion matrices of the primary emotion label that AMT subjects give for each video, contrasted with the ground-truth visual emotion (above) and audio emotion (below) in Task 2.}
\label{fig:amt_modality}
\end{figure}

\textbf{Task 2- Emotion Perception of Videos with Mismatched Emotions.}
In Task 2, subjects were asked the primary and secondary (if any) emotions they perceived from each video generated by our system, to investigate which modality people primarily rely on for emotion recognition. Overall, 426 of the 840 videos received two emotion labels.
We first compared the primary emotion label with the visual emotion (i.e., the condition emotion when generating the video) and the audio emotion, respectively. The confusion matrices are shown in Figure~\ref{fig:amt_modality}. Overall, 35.2\% of the primary emotion labels match with the visual emotion, while only 25.1\% of them match with audio emotion. If we only consider videos with mismatched emotions, these numbers become 31.4\% and 19.6\%, respectively. This suggests that the subjects relied on the visual modality much more heavily than the audio modality for emotion perception. Among the six emotions, happiness and disgust seem to be the easiest to perceive from the visual modality, while anger and fear are the most difficult. 

We then considered both primary and secondary emotions when comparing them with the audio and visual emotions. In this case, 44.9\% of labeled emotions can be matched to the visual emotion, while 33.8\% can be matched to the audio emotion. Similarly, if we only consider videos with mismatched emotions, these numbers become 41.1\% and 28.2\%. Again, this shows that the visual modality has a much greater effect than the audio modality on audiovisual speech emotion recognition for humans.

\section{Conclusions}
\label{sec:conc}
In this work, we proposed a novel emotional talking face generation system that is conditioned on speech, reference image, and categorical emotion inputs. We evaluated our network against the ground-truth videos and a baseline system ~\cite{vougioukas2019end} and validated that our method can generate emotional expressions effectively. In addition, we conducted a subjective study on AMT, showing that our method yields close performance to the ground-truth videos in terms of realness and emotion classification. Furthermore, we also conducted a pilot study on human emotion perception from audiovisual speech with mismatched emotions expressed in the audio and visual modalities, showing that visual perception is more dominant than auditory perception. For future work, we plan to improve the image quality of generated videos. We also plan to extend this work to 3D animation and rendering.

%



\section*{Acknowledgment}
This work is funded by the National Science Foundation (NSF) grant No. 1741472. You Zhang would like to thank the synergistic activities provided by the NRT program on AR/VR funded by NSF grant No. 1922591.


\ifCLASSOPTIONcaptionsoff
  \newpage
\fi



\bibliographystyle{IEEEtran}
\bibliography{main}
%



%

\begin{IEEEbiography}[{\includegraphics[width=1in,height=1.25in,clip,keepaspectratio]{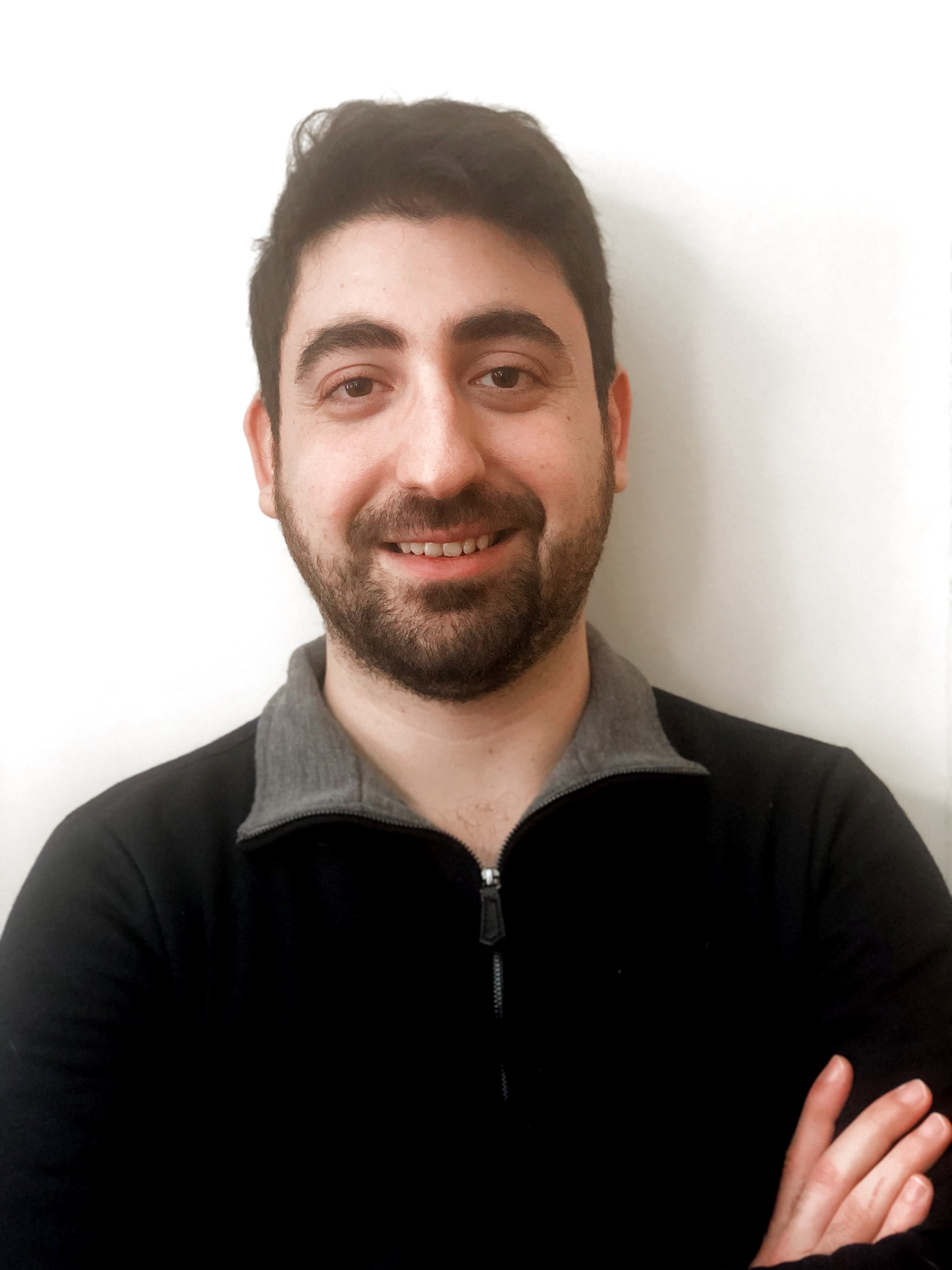}}]{S. Emre Eskimez} attended Sabanci University and graduated with a Bachelor of Science degree in Mechatronics Engineering in 2011. He began graduate studies in the Department of Mechatronics Engineering at Sabanci University in 2011 and received a Master of Science degree in 2013. He began graduate studies in the Department of Electrical and Computer Engineering at the University of Rochester in 2014, received a Master of Science degree in 2015, and received his Ph.D. in 2019. He joined Microsoft Cognitive Services Research Team (previously Speech and Dialog Research Group (SDRG)) in 2019. His research interests include speech enhancement, generative models, speech processing, natural language processing, multi-modal learning, and deep learning.
\end{IEEEbiography}

\begin{IEEEbiography}[{\includegraphics[width=1in,height=1.25in,clip,keepaspectratio]{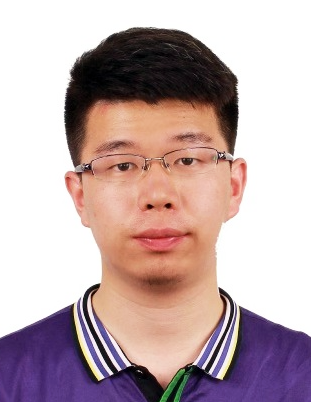}}]{You Zhang} received a B.E. degree in automation from the University of Electronic Science and Technology of China (UESTC), Chengdu, Sichuan, China, in 2019, and an M.S. degree in electrical engineering from the University of Rochester, Rochester, NY, USA, in 2021. He is currently a Ph.D. student in the Audio Information Research lab at the University of Rochester. His research interests lie in machine learning and its applications in speech and audio, such as audio-visual analysis, synthetic voice spoofing detection, spatial audio, etc.
\end{IEEEbiography}

\begin{IEEEbiography}[{\includegraphics[width=1in,height=1.25in,clip,keepaspectratio]{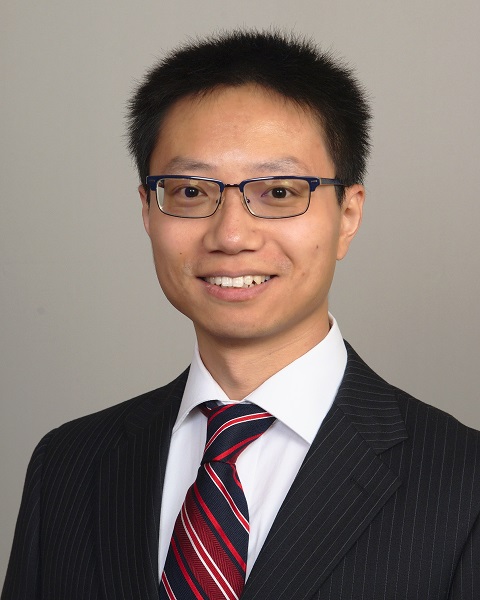}}]{Zhiyao Duan} (S’09, M’13) is an associate professor in Electrical and Computer Engineering, Computer Science and Data Science at the University of Rochester. He received his B.S. in Automation and M.S. in Control Science and Engineering from Tsinghua University, China, in 2004 and 2008, respectively, and received his Ph.D. in Computer Science from Northwestern University in 2013. His research interest is in the broad area of computer audition, i.e., designing computational systems that are capable of understanding sounds, including music, speech, and environmental sounds. He is also interested in the connections between computer audition and computer vision, natural language processing, and augmented and virtual reality. He received a best paper award at the 2017 Sound and Music Computing (SMC) conference, a best paper nomination at the 2017 International Society for Music Information Retrieval (ISMIR) conference, a BIGDATA award and a CAREER award from the National Science Foundation (NSF). His research is funded by NSF, NIH, and University of Rochester internal awards on AR/VR and health analytics.
\end{IEEEbiography}








\end{document}